\documentclass[%
	prb%
	,twocolumn%
	,a4paper%
	,showpacs%
    ]{revtex4}

\usepackage{graphicx}
\usepackage{upgreek}
\usepackage{xspace}
\usepackage[usenames,dvipsnames]{color}

\begin{document}

\newcommand{\etal}{\textit{et~al.}\xspace}
\newcommand{\degc}{\thinspace $\mathrm{^\circ C}$\xspace}
\newcommand{\mb}{$\upmu_{\mathrm{\scriptscriptstyle B}}$\xspace}

\title{Tracking defect-induced ferromagnetism in GaN:Gd}

\author{Martin Roever}
\author{Joerg Malindretos}\email[]{malindretos@ph4.physik.uni-goettingen.de}
\author{Amilcar Bedoya-Pinto}
\author{Angela Rizzi}
\affiliation{IV. Physikalisches Institut, Georg-August-Universt\"at G\"ottingen, Germany}
\author{Christian Rauch}
\author{Filip Tuomisto}
\affiliation{Department of Applied Physics, Aalto University, Finland}

\begin{abstract}
We report on the magnetic properties of GaN:Gd layers grown by molecular beam epitaxy (MBE). A poor reproducibility with respect to the magnetic properties is found in these samples. Our results show strong indications that defects with a concentration of the order of 10$^{19}$\thinspace cm$^{-3}$ might play an important role for the magnetic properties. Positron annihilation spectroscopy does not support a direct connection between the ferromagnetism and the Ga vacancy in GaN:Gd. Oxygen co-doping of GaN:Gd promotes ferromagnetism at room temperature and points to a role of oxygen for mediating ferromagnetic interactions in Gd doped GaN. 
\end{abstract}


\maketitle


The doping of non-magnetic semiconductors by magnetic metal ions (3d or 4f) has been among the hot topics of the last decade in the field of semiconductor spintronics. The aim is to introduce magnetic impurities that provide local spins in a non-magnetic host and to have a long range interaction mechanism between them that can lead to ferromagnetism at or above room temperature. Interestingly, also localized defect states can provide extra magnetic moments which might interact either with the magnetic ions or even just among themeselves~\cite{Dev:2008p117204,Gohda:2008p161201,Liu:2008p240,Mitra:2009p081202,Dev:2010p085207}. In these models the distances between the magnetic moments for a non-vanishing exchange interaction are usually below about 10\thinspace $\text{\AA}$. In this case the percolation threshold that leads to the onset of a macroscopic ferromagnetic state requires concentrations of localized moments of about 1\thinspace\%~\cite{OsorioGuillen:2007p184421}. Such high concentrations can usually not be obtained in thermal equilibrium and therefore the experiments require a high level of control and the results must be critically analyzed. Theory has promoted a rush for room temperature ferromagnetism in the field of dilute magnetic semiconductors. Important aspects and results have been recently reconsidered and critically reviewed by Zunger~\etal\cite{Zunger:2010p53}.

The first report on ferromagnetic GaN:Gd was on a MBE-grown layer with about 6\thinspace\% of Gd~\cite{Teraguchi:2002p651}. Exceptionally high effective magnetic moments of up to 4000\thinspace\mb per Gd atom were found in highly dilute GaN:Gd layers~\cite{Dhar:2005p037205,Dhar:2005p245203}. Later it was shown that these moments can even be increased to 10$^5$\thinspace\mb in Gd ion implanted GaN layers~\cite{Dhar:2006p062503}. Nearly all recent reports about Gd doped GaN show a strong ferromagnetic signal at room temperature~\cite{Teraguchi:2002p651,Dhar:2005p037205,Dhar:2005p245203,Roever:2008p2352,Zhou:2008p062505,Davies:2010p212502}. Only few exceptions show partly ferromagnetic samples or no ferromagnetism~\cite{Hite:2007p391,Lo:2007p262505}. 

The origin of the ferromagnetic order in GaN:Gd is still unknown. It was shown that cation vacancies can provide local moments which exchange directly by their overlapping wavefunctions and provide a long ranged coupling mechanism~\cite{Dev:2008p117204,Gohda:2008p161201}. According to Neugebauer \etal the cation vacancy in GaN is a rather unlikely defect if the Fermi energy is not close to the conduction band minimum~\cite{Neugebauer1994}. Furthermore, nitrogen or oxygen interstitials in GaN:Gd can also provide local moments with a long ranged coupling and their formation is much more likely than the cation vacancy~\cite{Mitra:2009p081202}. 

In this paper we show experimental results about the reproducibility and long term stability of the ferromagnetic phase in Gd doped GaN layers. The theory of the Ga vacancy model is probed experimentally by positron annihilation spectroscopy (PAS). Oxygen co-doping of GaN:Gd was carried out to introduce oxygen impurities on purpose.


All GaN:Gd samples were grown by plasma assisted molecular beam epitaxy on metal organic chemical vapor deposition (MOCVD) GaN/Al$_2$O$_3$(0001) templates. If not stated otherwise, the MBE process was carried out with our optimized parameters for the growth of unintentionally doped GaN layers, namely 760\degc substrate temperature and slightly metal rich conditions. Gd (4N) was supplied during growth by a high temperature effusion cell. Oxygen co-doping was carried out applying oxygen gas (4.8N) with a flux ranging from 0.1 to 0.5\thinspace sccm into the growth chamber. The growth time was kept constant for all epitaxial layers, resulting in a thickness of about 500\thinspace nm for the standard GaN:Gd layers and 200\thinspace nm for the oxygen co-doped samples.

A superconducting quantum interference device (SQUID) MPMS 5 from Quantum Design was used to measure the magnetic properties. All data have been corrected for diamagnetic background and trapped field artifacts \cite{Qdesign:2002n1014-213,QDesign:2001n1014-208}. The resolution limit is about $3 \cdot 10^{-7}$\thinspace emu, which translates to about $5 \cdot 10^{18}$\thinspace\mb/cm$^3$ for our sample geometry. Reference measurements on the substrate material were made on a regular basis and no ferromagnetic contributions were found.

The Gd concentration is measured by time of flight secondary ion mass spectroscopy. The resolution limit of Gd isotopes is about $10^{16}$\thinspace cm$^{-3}$. All lower concentrations are extrapolated by effusion theory. The contamination with transition metals can be estimated to be at least a factor of 10$^3$ lower than the Gd concentration~\cite{BedoyaPinto:2009p195208}.

The Doppler broadening of the positron annihilation radiation was measured by two germanium detectors with a resolution of $1.3$\thinspace keV at $511$\thinspace keV, using a mono-energetic slow positron beam. A spectrum of $n > 5 \cdot 10^{5}$\thinspace counts was accumulated for each point. The lineshape of the Doppler broadened annihilation $\upgamma$\hbox{-}radiation is analyzed using the conventional lineshape parameters S ($\mathrm{|E_{\upgamma}-511\thinspace keV| < 0.75}$\thinspace keV) and W (2.86\thinspace keV $\mathrm{< |E_{\gamma}-511\thinspace keV| < }$ 7.33\thinspace keV). All points have been normalized to the value of a suitable reference sample where positrons annihilate solely in the delocalized state in the GaN lattice.


The structural quality of the Gd doped epitaxial layers below a Gd concentration of $10^{20}$\thinspace cm$^{-3}$, at which GdN clusters are formed, resembles those of unintentionally doped GaN, as deduced from X-ray diffraction analysis as well as transmission electron microscopy (not shown). X-ray absorption near edge structure (XANES) analysis revealed that the Gd is incorporated substitutional on Ga sites in a 3+ valence state and no significant bond length variations of the \mbox{Ga-N bond} was found~\cite{MartinezCriado:2008p3198}.

\begin{figure}
\center
    \includegraphics[width=.9\columnwidth]{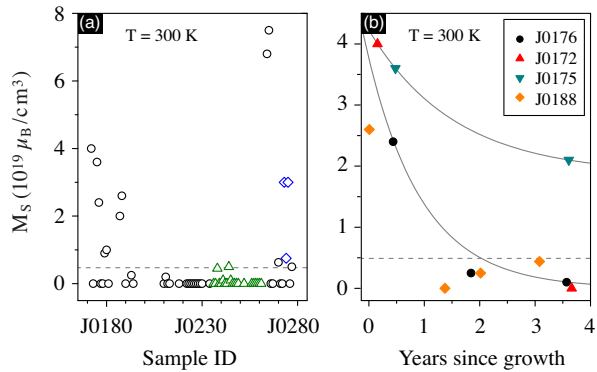}
    \caption{(a) Saturation magnetization at 300\thinspace K against the sample ID. Black circles show Gd doped GaN samples grown under optimized conditions. Green triangles represent samples grown under variing conditions as explained in the text. Blue diamonds stand for oxygen co-doped GaN:Gd. (b) Saturation magnetization against time for selected samples. The dotted line is the resolution limit of the SQUID magnetometer and the exponential curves are shown as a guide to the eye.}
    \label{fig:saturation}
\end{figure}%
\begin{figure}
\center
    \includegraphics[width=.85\columnwidth]{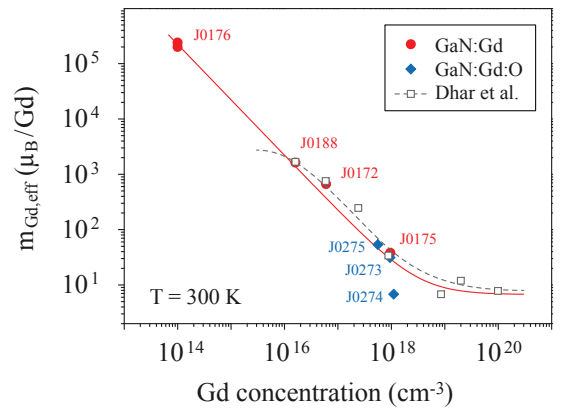}
    \caption{Effective saturation magnetization of ferromagnetic GaN:Gd samples against Gd concentration. Red dots and blue diamonds are samples of this work, for comparison the data from Dhar \etal is shown as open squares~\cite{Dhar:2005p245203}. The red line represents a fit as discussed in the text.}
    \label{fig:saturation_conc}
\end{figure}%

In Fig.~\ref{fig:saturation}(a) the saturation magnetization M$_\mathrm{s}$ at 300\thinspace K measured for many GaN:Gd epitaxial layers is plotted against the sample ID, i.e. in chronological order. In all samples the Gd concentration is in the range of 10$^{14}$ to 10$^{19}$\thinspace cm$^{-3}$. The series around sample number J0180 clearly shows sizable M$_\mathrm{s}$ values~\cite{Roever:2008p2352}. On the other hand several samples that have been grown later with the same nominal growth parameters (black circles) show a vanishing saturation magnetization at room temperature, with the exception of J0264 and J0265. The green triangles correspond to samples grown in a wide range of III-V ratios, slightly different substrate temperatures and within the same range of Gd concentrations, but no influence of these parameters on the magnetic properties was found. Oxygen co-doping of three samples (blue diamonds) was found to induce a ferromagnetic order at room temperature with M$_\mathrm{s}$ values comparable to that of the first series. The saturation magnetization value increases with the oxygen supply during growth. It should be noted, that GaN:Gd samples without oxygen co-doping grown directly before and after the co-doped samples exhibit no ferromagnetism. Selected samples have been repeatedly characterized by SQUID. The saturation magnetization is observed to decrease with time (Fig.~\ref{fig:saturation}(b)).

A first general result from these experiments is a very poor reproducibility of the magnetic behavior. Thus, uncontrolled parameters seem to be responsible for the ferromagnetic properties in the GaN:Gd epitaxial layers grown by MBE. In particular the role played by either intrinsic or extrinsic point defects has to be taken into account. It is worth noting that the presumed defects responsible for the ferromagnetic properties are only induced if Gd is supplied and incorporated into the layers during growth, since none of our undoped GaN layers shows ferromagnetism. 

Ferromagnetic and non-ferromagnetic samples have been characterized by photoluminescence at 2\thinspace K and the main observation is a clear enhancement of the defect related luminescence for the ferromagnetic GaN:Gd layers (not shown), but no clear assignment is possible. The role of defects was also pointed out in experiments with Gd-implanted GaN and as-grown GaN:Gd epitaxial layers~\cite{Dhar:2006p062503,Davies:2010p212502,Mishra2010}.

The dependence of the effective saturation magnetization per Gd atom $m_\mathrm{Gd,eff}$ on the Gd concentration $c_\mathrm{Gd}$ is depicted in Fig.~\ref{fig:saturation_conc} and compares well with values in the literature~\cite{Dhar:2005p245203}. The observed trend could be described by a constant additional magnetic moment $M_\mathrm{{def}} = 2.5 \cdot 10^{19}$\thinspace \mb/cm$^{3}$ independent of the Gd concentration, according to the expression $m_\mathrm{Gd,eff} = m_\mathrm{Gd} + M_\mathrm{def} / c_\mathrm{Gd}$. Assuming a defect driven mechanism of the ferromagnetism, an effective defect concentration of this order would be required. In this scenario colossal magnetic moments could be explained with a high but not unrealistic concentration of magnetically active defects. However, it is unclear why these defects only form in the presence of Gd while their amount would be independent of the Gd concentration. 

Electrical-transport properties of GaN:Gd grown on highly resistive 6H-SiC(0001) substrates show that even Gd concentrations lower than the background carrier concentration of typically $10^{17}$\thinspace cm$^{-3}$ increase the resistivity by several orders of magnitude as compared to unintentionally doped GaN~\cite{BedoyaPinto:2009p195208}. The observed variable-range-hopping in an impurity band of localized states is another hint, that deep defects form in a large number, even for low Gd concentrations.

Secondary ion mass spectroscopy (not shown) reveals oxygen concentrations above $5 \times 10^{18}$~cm$^{-3}$ in our GaN:Gd layers (J0175, J0188) and clearly higher values for the oxygen co-doped samples (J0273 -- J0275). The high resistivity of the GaN:Gd layers indicates that the oxygen atoms are not mainly incorporated substitutionally on nitrogen sites, where they would form shallow donors.

\begin{figure}
\center
  \includegraphics[width=.8\columnwidth]{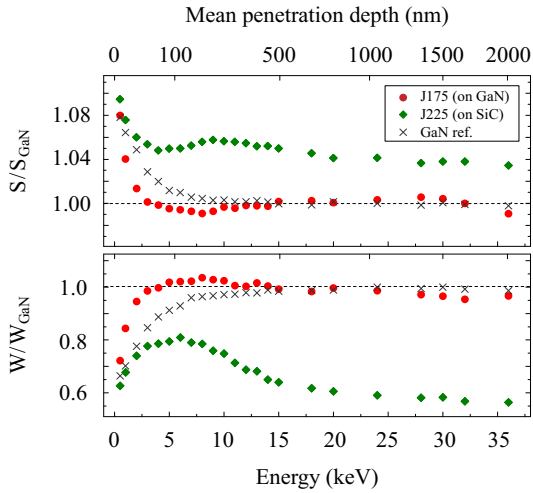}
   \caption{S and W parameter for samples J0175 (magnetic) and J0225 (non-magnetic) as a function of the positron implantation energy. All spectra are divided by the GaN reference value (see crossed spectrum for comparison).}
   \label{fig:sw_e}
\end{figure}

\begin{figure}
\center
  \includegraphics[width=.85\columnwidth]{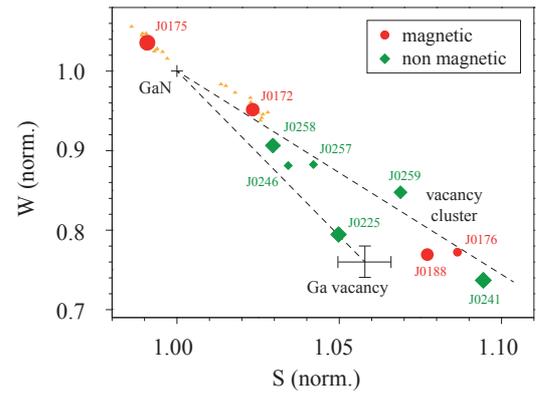}
	\caption{SW-plot for the layer specific lineshape parameters of selected samples at room temperature. Results from temperature dependent measurements are marked with small triangles. The point size scales with the Gd concentration. Characteristic values for the Ga vacancy~\cite{Tuomisto2007} and vacancy clusters~\cite{Rummukainen2004} are included.}
   \label{fig:sw_plot}
\end{figure}

With the aim of investigating possible correlations between ferromagnetism in GaN:Gd and the presence of cation vacancies we performed positron annihilation spectroscopy (PAS). Depth dependent Doppler broadening spectra have been recorded at room temperature (Fig.~\ref{fig:sw_e}). From these, the S and W lineshape parameters representative for the GaN:Gd layers are determined and displayed in Fig.~\ref{fig:sw_plot}, together with characteristic values for the GaN lattice and the gallium vacancy~\cite{Tuomisto2007}. In order to ensure that the determined parameters at room temperature are representative for the vacancy concentration in the samples and not influenced by positrons detrapping from vacancies to negative ions~\cite{Saarinen1997}, temperature dependent measurements up to 550\thinspace K have been performed for selected samples (Fig.~\ref{fig:sw_plot}, small triangles). The reference value for the GaN lattice is determined by measuring a reference sample in which no positron trapping to vacancy defects is observed (Fig.~\ref{fig:sw_e}).

The identity of dominant positron traps in a certain sample can be evaluated based on its position in the SW-plot. The latter is given as the linear combination of the characteristic values of the involved annihilation states, weighted with the fraction of positrons annihilating in these states. The measured SW points for the GaN:Gd layers indicate a broad defect landscape in the different samples which is dominated by trapping to vacancy clusters. The strong deviation from the Ga vacancy line and high relative S parameters of up to 1.09 are characteristic for vacancy clusters with a considerably larger open volume than the isolated Ga vacancy. The relevant vacancy cluster point is not known exactly, but similar S parameters have been observed before in MBE grown GaN~\cite{Rummukainen2004}.
We can estimate the vacancy cluster density to lie in the range between $1 \times 10^{16}$~cm$^{-3}$ and $1 \times 10^{18}$~cm$^{-3}$. Since the accurate SW-reference point for these defects is not known, the upper bound could even be lower by about one order of magnitude.

Considering only the ferromagnetic samples in Fig.~\ref{fig:sw_plot} (red circles), the strongest cluster signal is found for the sample with the lowest Gd-concentration (J0176, $\mathrm{c_{Gd} = 1.5 \cdot 10^{14}}$\thinspace cm$^{-3}$) and the cluster signal decreases with increasing Gd doping. In the sample with the most stable ferromagnetism with time and highest absolute saturation magnetization (J0175, $\mathrm{c_{Gd} = 9.6 \cdot 10^{17}}$\thinspace cm$^{-3}$), the measured SW point is located close to the GaN lattice and hence no positron trapping to vacancy defects is observed. At room temperature the positron diffusion length in this sample is considerably reduced compared to the GaN reference (Fig.~\ref{fig:sw_e}). This hints at the presence of a high concentration of negatively charged ions~\cite{Saarinen1997}, which can trap positrons through the formation of Rydberg states and possess annihilation characteristics close to the GaN lattice. It should be noted that detrapping from vacancy defects could be explicitly excluded in this sample, due to the temperature dependent measurements. For the non-magnetic samples the PAS results do not evidence any clear correlation between the Gd concentration and the vacancy type or concentration. Sample J0225, whose Gd content of $\mathrm{5.4 \cdot 10^{17}}$\thinspace cm$^{-3}$ is comparable to that of J0175, lies close to the saturation point for single Ga vacancy related defects. Nevertheless, this sample never showed any ferromagnetic behavior. So, the defect landscape that has been probed by PAS, namely single gallium vacancies and larger open volume defects, does not directly correlate with the magnetic properties of the GaN:Gd epitaxial layers. We can therefore conclude that the cation vacancy models proposed in literature are not supported by our experiments \cite{Dev:2008p117204,Gohda:2008p161201,Liu:2008p240}.  

Based on formation energy arguments, interstitial nitrogen as well as oxygen in octahedral sites next to Gd have been proposed to be a more likely source of defect induced magnetism in GaN:Gd than cation vacancies~\cite{Mitra:2009p081202}. Both types of interstitials induce a spin split band of localized states originating from the non-bonding N\mbox{-}2p or O\mbox{-}2p orbitals, which lies well within the energy gap. Both configurations would be consistent with the observed compensation of the unintentional n-type dopants in GaN:Gd and with the electrical transport by hopping \cite{BedoyaPinto:2009p195208}. Due to the large affinity of Gd to O it is also possible that oxygen donors are attracted into interstitial positions by the Gd, thus further contributing to the compensation and providing extra localized magnetic moments. The efficient gettering of residual donors in GaN by Gd was also reported for bulk samples grown under high pressure~\cite{Lipinska:2006p243} and was theoretically confirmed by Mitra and Lambrecht for the case of substitutional oxygen~\cite{Mitra:2009p081202}. 

The non-reproducibility of the ferromagnetic properties in our samples could be explained by an uncontrolled balance between the native donors and the compensating defects. If the concentration of localized magnetic moments were near the percolation threshold, small variations of their amount could switch the macroscopic ferromagnetic state, as discussed in the appendix of Osorio-Guillen~\etal~\cite{OsorioGuillen:2007p184421}. In this case the favorable effect of co-doping with oxygen during growth, as seen for the samples in Fig.~\ref{fig:saturation}~(blue diamonds) might be explained by an increase of interstitial oxygen in presence of Gd. Therefore the model of a ferromagnetic state induced by interstitial oxygen is at least qualitatively consistent with our data. It should be noted, that attributing the favourable effect of oxygen co-doping to carrier induced ferromagnetism due to O$_\mathrm{N}$ donors seems rather unlikely considering the high resistivity of ferromagnetic GaN:Gd~\cite{BedoyaPinto:2009p195208,Dhar:2005p037205}.


In conclusion, room temperature ferromagnetism and colossal magnetic moments of dilute GaN:Gd layers grown by MBE are confirmed. However, the reproducibility and the long term stability of the magnetic properties are poor. Uncontrolled parameters seem to be responsible for the ferromagnetic properties. It is shown that a concentration of magnetically active defects of about 10$^{19}$\thinspace cm$^{-3}$ might explain the observed magnetic properties. PAS measurements rule out single gallium vacancies as the origin of magnetic coupling in as-grown GaN:Gd and do not show any direct connection of gallium vacancy clusters to room temperature ferromagnetism in this material either. Oxygen co-doping is observed to promote ferromagnetism in GaN:Gd. 
However, more investigations are necessary to indubitably identify the role of oxygen and the magnetic coupling mechanism in this material.

\begin{acknowledgments}
We thank Andreas Laufer (Justus-Liebig-Universit\"at Gie\ss en) for the SIMS measurements and the Deutsche Forschungs Gesellschaft for funding within the SFB\thinspace 602.
\end{acknowledgments}


\end{document}